# Visible Light Triggered Drug Release from $TiO_2$ Nanotube Arrays: A Novel Controllable Antibacterial Platform

Jingwen Xu, [a] Xuemei Zhou, [b] Zhida Gao, [a] Yan-Yan Song, [a]* and Patrik Schmuki[b]*

---

[a] J. Xu, Dr. Z. Gao, Prof.Dr. Y.-Y. Song, College of Sciences, Northeastern University, Shenyang 110004, China, E-mail: yysong@mail.neu.edu.cn

[b] X. Zhou, Prof. Dr. P. Schmuki, Department of Materials Science, WW4-LKO, University of Erlangen-Nuremberg, Martensstrasse 7, D-91058 Erlangen, Germany, Email: schmuki@ww.uni-erlangen.de

**Abstract:** In this work, we use a double-layered stack of $TiO_2$ nanotubes (TiNTs) to construct a visible-light triggered drug delivery system. Key for visible-light drug release is a hydrophobic cap on the nanotubes containing Au nanoparticles (AuNPs). The AuNPs allow for a photocatalytic scission of the hydrophobic chain under visible light. To demonstrate the principle, we loaded antibiotic (ampicillin sodium (AMP)) in the lower part of the $TiO_2$ nanotube stack, triggered visible light induced release, and carried out antibacterial studies. The release from the platform becomes most controllable if the drug is silane-grafted in hydrophilic bottom layer for drug storage. Thus visible-light photocatalysis can also determine the release kinetics of the active drug from the nanotube wall.

Controllable drug delivery has attracted wide interest in biomedicine and other fields of science in order to achieve a targeted use of an active substance at the right time and the right place. Particularly drug release mechanisms with triggers that respond to surrounding factors, such as pH,[1] temperature,[2] illumination and ionic strength[3] are of a high practical significance. Many nanomaterials are, or can be designed to be particularly sensitive to environmental factors and therefore a steeply increasing number of studies is being carried out on the development of "smart" nanomaterial-based drug carriers. Systems have been developed with great biocompatibility and the feasibility of targeted drug delivery with a much higher control over the pharmacokinetics (which can decrease systemic toxicity).[4, 5] For example, Dai and his coworkers prepared pH-sensitive polyethylene glycol functionalized graphene oxide for the delivery of aromatic drugs,[6a, 6b] or hierarchical hollow $CaCO_3$ nanoparticles have been reported for a localizing drug release reacting on a pH change.[6c] Another example is the work of Zhang et al. that reported a free-radical precipitation polymerization method to synthesize a temperature responding drug carrier.[7]

As drug carrier a wide range of nanoparticles,[8] microgel,[9] nanotubes[10, 11] and polymeric micelles[12, 2b] have been explored. Especially, drug carriers based on nanotubes have various beneficial features due to their intrinsic high surface-to-volume ratio, well defined geometry and stable structure.[13] In the past decade $TiO_2$ nanotube arrays (TiNTs) grown by a self-organizing electrochemical anodization process have attracted tremendous scientific interest due to the combination of geometric features with an inherent photocatalytic activity. TiNTs have shown great potential in optics, energy storage, bioelectronics as well as for medical devices.[14] Biocompatible scaffolds for hosting functional guest molecules can be fabricated, taking advantage of the large number of hydroxyl groups present on the tube walls. These provide the possibility of incorporating desired functional groups, to load drugs or graft capping molecules and thus establish advanced storage and release mechanisms. For example, recently, our group has reported the combination of TiNTs for payload loading and controllable release by UV light irradiation.[15] However, UV light only accounts for 2~4% of solar energy and thus such a trigger mechanism can't respond well to solar light. Even more important, many biomolecules suffer from denaturation or disintegrate when long-term irradiated by UV light.

In the present work, we introduce a visible-light triggered drug delivery platform where the trigger for releasing the drug is based on Au-SPR induced photocatalytic chain scission of a hydrophobic cap on the nanotubes. As an example we demonstrate a highly controllable antibacterial activity, by releasing under visible light, an antibiotic drug (AMP) loaded into these amphiphilic tubes.

Scheme 1 illustrates the steps used for the preparation of this TiNTs based platform. It consists of two parts: the top hydrophobic layer that acts as a cap and the lower hydrophilic layer serves for antibacterial drug storage. For synthesis (see also details in Figure S1-S3), first, a $TiO_2$ nanotube layer is grown in a glycerol/water/$NH_4F$ electrolyte to a thickness of ~0.3 μm with individual nanotube diameters of approx. 90 nm (Figure 1A and Figure 1B). The as-formed amorphous nanotubes are then crystallized to anatase by annealing at 450 $^oC$ in air for 1 h (anatase provides a much higher photocatalytic activity than amorphous material). AuNPs are decorated on the tube wall and entrance (Figure 1C and 1D) by a biotemplated method that we reported before (and is briefly outlined in Figure S1).[16] After Au decoration, a hydrophobic monolayer of octadecylphosphonic acid (ODPA) is attached to the tube walls (the successful decoration of AuNPs and ODPA was characterized by XPS as plotted in Figure S3). The sample is subsequently anodized again in an ethylene glycol/$NH_4F$ electrolyte. In contrast to water-based electrolytes, ethylene glycol electrolytes enter into the hydrophobic tubes and therefore allow for a second anodic tube growth through the bottom of the first tube layer. The voltages for second anodization were chosen to match the nanotube diameter in the first layer (30 V). The length of the lower layers is controlled by the duration of the second anodization. As shown in Figure 1B, the side view verifies that nanotubes can grow aligned to the upper layer during the second anodization - the lengths of the second layer used here were about 1.7 μm determined by the duration of the second anodization of 3 h.

In principle, to load a higher amount of drug, a longer anodization time to grow a thicker second layer would be advantageous (Figure S4). However, there is a detrimental influence on the hydrophobic properties if the ODPA layer is exposed to a high voltage (electric field) for an extended time.[17] Therefore, a time of 3 h represents an optimized duration for the second anodization (see results in Figure S5).

After the growth of the lower layer, dodecanethiol (NDM) was used to additionally coat the defects formed in ODPA hydrophobic layer during the second anodization in order to strengthen the overall hydrophobic nature of the cap layer. To load the lower part of the tubes with the active drug (AMP), first (3-glycidyloxypropyl) trimethoxysilane (GPMS) molecules were attached to the walls of the lower part of the nanotubes then the samples were immersed into an ethanol solution of AMP. AMP molecules react with the silane linker to form a covalent bond as outlined in Figure S2[18].

In order to characterize the effect of the different preparation steps on the surface wettability we measured water droplet contact angles ($\theta\omega$) after critical synthesis steps (Figure 2). The annealed TiNTs and TiNTs/AuNPs show a completely hydrophilic wetting characteristic ($\theta\omega \approx 0$). After incubation with ODPA that attaches to the wall of nanotubes, the surface of the sample shows a strongly hydrophobic character with $\theta\omega = 124.6^o$ (Figure 2A). After the second anodization, the contact angle exhibits a slight decrease, but the hydrophobic outer nature is maintained with $\theta\omega = 104.2^o$ (Figure S3). This demonstrates that most of the ODPA molecules in the first layer can largely "withstand" the second anodization step in the fluoride-containing ethylene glycol electrolyte. The second incubation in dodecanethiol (NDM) – that decorates the Au nanoparticles – leads again to an increase of contact angle to $\theta\omega = 141.3^o$ (Figure 2B). Only a slight decrease of the contact angle ($\theta\omega = 132.8^o$) can be observed after attachment of the silane and AMP to the walls of the lower nanotubes. This illustrates the strongly hydrophobic character of the tube layer surface after the entire synthesis of the drug loaded platform. These processes are characterized by XPS as shown in Figure S3.

To evaluate visible-light activity introduced by the Au nanoparticle loading and successful electronic coupling of Au with the $TiO_2$ tubes, we acquired not only reflectivity data (Figure 3A), but also photocurrent spectra (Figure 3B). The light absorption data show an Au nanoparticle induced absorption behavior in the visible range peaking at ~526 nm (curve b, Figure 3A) due to surface plasmon resonance (SPR).[19, 20] The photograph (inset of Figure 3A) of the amphiphilic $TiO_2$ nanotube arrays after Au nanoparticle decoration also shows the pink color typical for the SPR of Au nanopaticles. The photocurrent spectra for the AuNPs containing sample show an onset that is red shifted and an enhancement (curve b, Figure 3B) over the entire spectral range. This broad enhancement can be explained by: once visible light absorption takes place a photocatalytic decomposition of the organic monolayer can start.[21] I.e. once the first hydrophobic organic chains are broken, wetting of the tube tops increases, providing better electrolyte access to the nanotube arrays, thus enhancing charge transfer and as a result the photocurrents.

In order to evaluate visible light activity under conditions without an applied bias (OCP) - as the platform would be used under operational conditions - we carried out a color reaction test as shown in Figure S6 to demonstrate the formation of $H_2O_2$. The positive reaction for Au decorated tubes (and the negative results on a non-decorated reference) strongly supports that Au-SPR can provide a conduction band induced photocatalytic pathway.[21a] In literature for such systems not only the excitation and flow of electrons from the Au deposits to the conduction band (CB) of $TiO_2$ has been reported,[21b] but the decorated Au nanoparticles can also serve as trapping centers for electrons photogenerated in the CB of $TiO_2$. Under visible light conditions this may lead to an improved photocatalytic efficiency for AuNPs decorated TiNTs mainly by conduction band induced $O_2^{\bullet}$ formation.[21] In the UV range photo-induced valence band holes can additionally contribute and break hydrophobic chains by $^{\bullet}OH$ formation or direct hole transfer.

To demonstrate that the present system indeed can show an effective and controllable release of a real drug under bias-free and visible-light-only conditions, AMP release kinetics were studied for different drug anchoring and capping stages (Figure 4). For this, samples were exposed to a Xe light source (with a filter of $\lambda$ > 420 nm, illumination intensity ~50 mW/cm$^2$) after immersion into DI water (5 mL) and keeping it in the dark for 10 min. To detect the released drug we used UV-vis spectrometry (after reaction with ninhydrin, Figure S7). As apparent from Figure 4A, if the AMP is only loaded by dipping (neither cap nor linker is used), a quick and uncontrolled release is obtained. Most of the drug molecules are released from the nanotubes in the first 10 min i.e. also in the dark period. With hydrophobic Au-caps, AMP is retained within the tubes in the dark but the drug molecules are released immediately after the removal of the upper cap by visible light (Figure 4B). Figure 4C shows the drug release kinetics for hydrophilic nanotubes, in which AMP is covalently linked with the tube wall by GPMS but no Au coating cap is present. In this case, under light there are almost no AMP molecules released due to a lack of photocatalytic activity. This demonstrates the visible light cannot trigger the release of AMP without grafting AuNPs on the tubes. Figure 4D shows the results from amphiphilic TiNTs where AuNPs are introduced into the hydrophobic cap and the AMP is covalently linked to the lower tube walls. These layers enable an illumination controlled release. In the first dark period, there is a very minor release that can be ascribed to small amount of physical adsorbed AMP still present. Upon illumination, however, hydrophobic ODPA molecules on the first nanotube layer are decomposed, the GMPS linkers are cut and AMP molecules are controllably released.

To proof full functionality of this platform, we investigated the release of AMP in a bacterial culture test. Escherichia coli (*E. coli*), which is responsible for many infections in daily life, served as the model target microorganism for antibacterial tests. In Figure 5A, the antibacterial activity from hydrophilic TiNTs (anatase tube without AMP loading) and empty amphiphilic AuNPs/TiNTs (without AMP loading) were compared with the amphiphilic AuNPs/TiNTs loaded AMP (glass slides were used as control experiments). Clearly, the amphiphilic nanotubes with drug loading show an effective bactericidal activity under visible light irradiation.

The hydrophilic nanotubes and the empty amphiphilic AuNPs/TiNTs exhibit only a weak bactericidal efficiency, suggesting the photocatalytic antibacterial activity of the nanotubes and direct visible light effects can be neglected. In addition, we evaluated the controllability of the drug delivery system by releasing the drug for different illumination periods (Figure 5B). The sample was first kept in dark during the first ten minutes then illumination started and bacterial efficiency was studied in 30 min time intervals. The results indicate a light controllable amount of drug release corresponding to the illumination duration. In these experiments the bactericidal effect tapers off after 90 min, suggesting that at this point the available drug molecules are completely released. The influence of visible light is also clearly evident in Figure 5C and 5D, where clearly a drastic decrease of *E. coli* colonies is observed using the amphiphilic drug release system under visible light irradiation (Figure 5C) if compared with the results obtained in dark (Figure 5D). The result demonstrates the hydrophobic cap can efficiently prevent the leaching of the AMP or the influx of the aqueous surrounding media in dark. I.e., the drug delivery system is comparably stable against drug leakage if stored in dark. XPS data in Figure S3 and Figure S8 also confirm the photocatalytic scission of the hydrophobic chain and the alkyl chain that connects the drug to wall by visible light irradiation. Moreover, if the Au decorated top layer is removed before visible light irradiation (Figure S9), the drug molecules cannot be released.

Another very important point evidenced from the results of Figure 5B is that the visible light induced photocatalytic effect allows the release of the antibiotic species in a fully functional form. That is, neither the used photocatalytic principle nor the direct visible light illumination, have a detrimental effect on the AMP molecules in their function – this in contrast to UV induced approaches.

In summary, we established a visible light controlled platform for drug delivery based on $TiO_2$ nanotubes. The key trigger mechanism is visible light photocatalytic chain scission of an organic monolayer. Activation for visible light is introduced by Au nanoparticles embedded in the hydrophobic cap layer, where Au-SPR with the $TiO_2$ conduction band provides the active species for chain scission. The system-functionality was tested in antibacterial experiments towards *E. coli*. It shows the highest degree of release-control when the drug is not only loaded but additionally anchored by a silane linker to the tube wall (i.e. using a second chain scission step to control release). The results, very importantly, show that the loaded drug can be photocatalytically released in its fully functional form (i.e. without photoinduced degradation). We believe the principle shown here, that is Au SPR/$TiO_2$ induced photocatalytic chain scission by visible light is not limited to the drug release system demonstrated but may be expanded to another payload release systems for much wider use.

**Acknowledgements**

This work was supported by the National Natural Science Foundation of China (No. 21322504, 11174046, 21275026), the Fundamental Research Funds for the Central Universities (N140505001, N140504006). we also acknowledge financial support from the Alexander von Humboldt Foundation (for Y.Y.S.). And we would also like to acknowledge ERC, DFG and DFG cluster of excellence for their financial support.

**Keywords:** surface plasmon resonance • drug delivery • nanotubes • antibiotics •

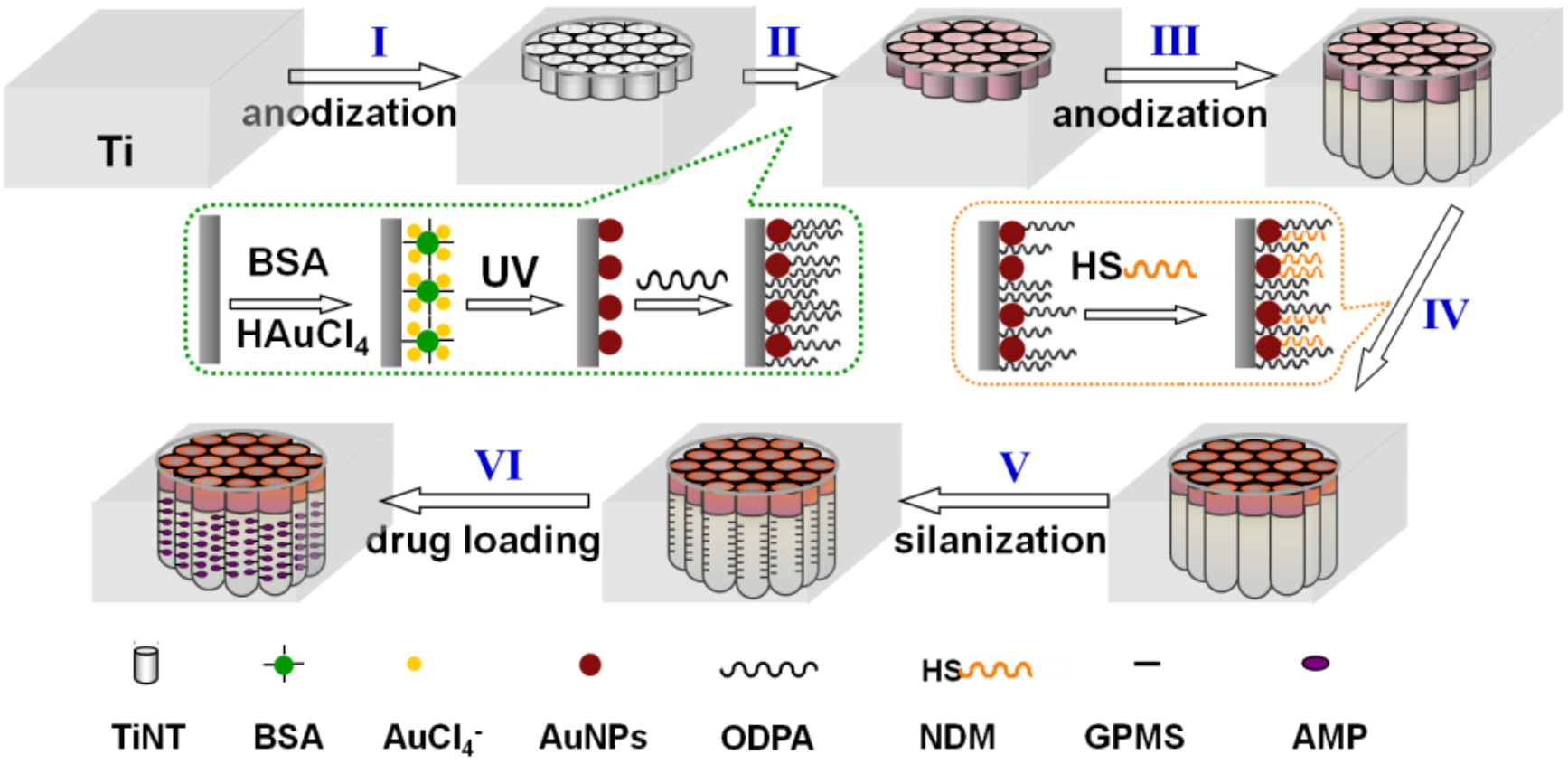

**Scheme 1.** Preparation of visible light controlled drug-release TiNTs. I: First anodization to form TiNTs; II: Decoration of Au nanoparticles and hydrophobic monolayer to TiNTs; III: Second anodization to allow nanotube continue to grow for drug storage; IV: Link of hydrophobic monolayer to Au nanoparticles by mercaptan; V: Silanization of lower layer with GPMS; VI: Drug (AMP) loaded by GPMS linker (the details are shown as Figure S2 in Supporting Information).

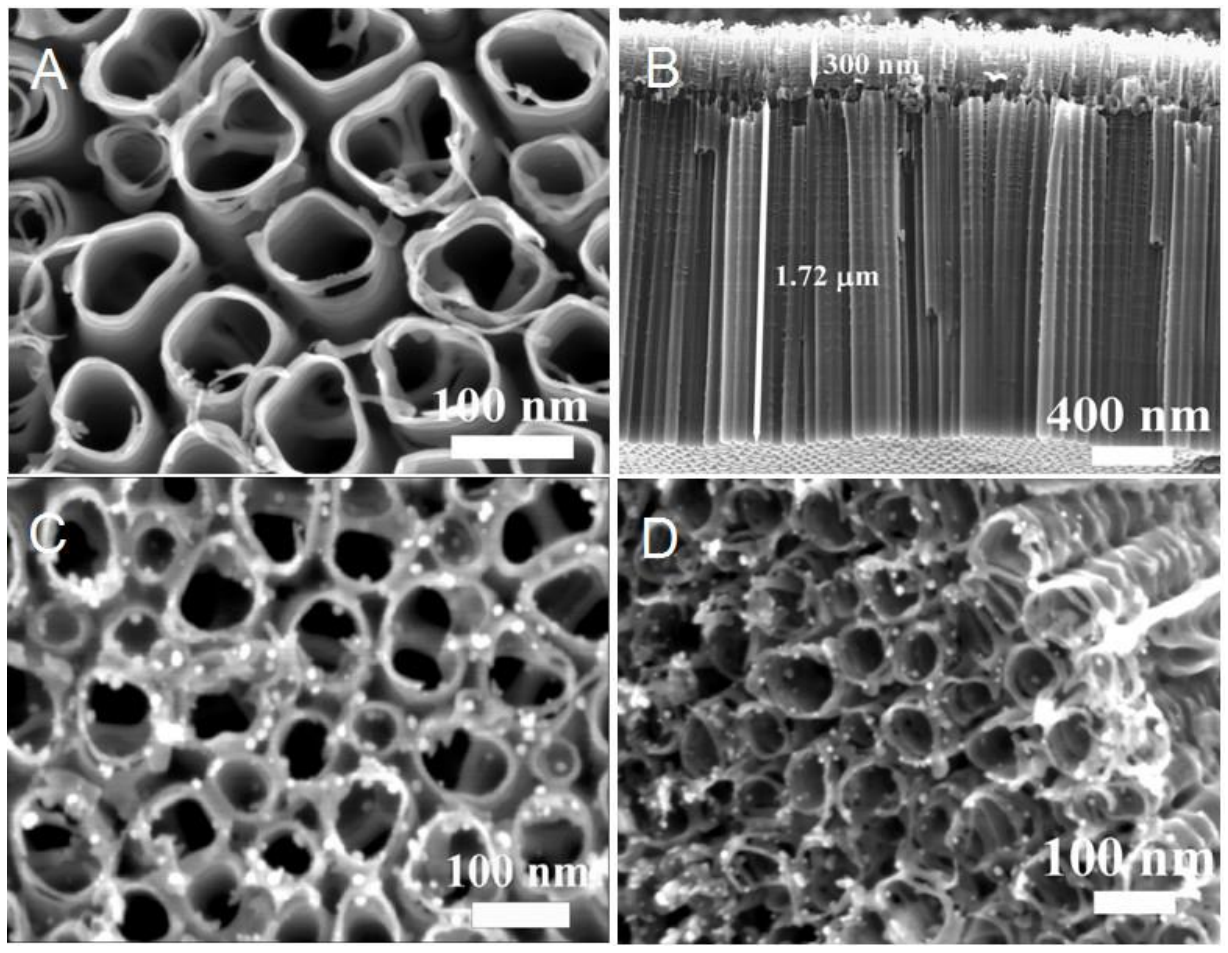

**Figure 1.**SEM images of (A) top-view images and (B) cross-sectional images for double nanotube layers TiNTs; (C) top-view and (D) cross-sectional images for upper nanotube layers of TiNTs/AuNPs-ODPA.

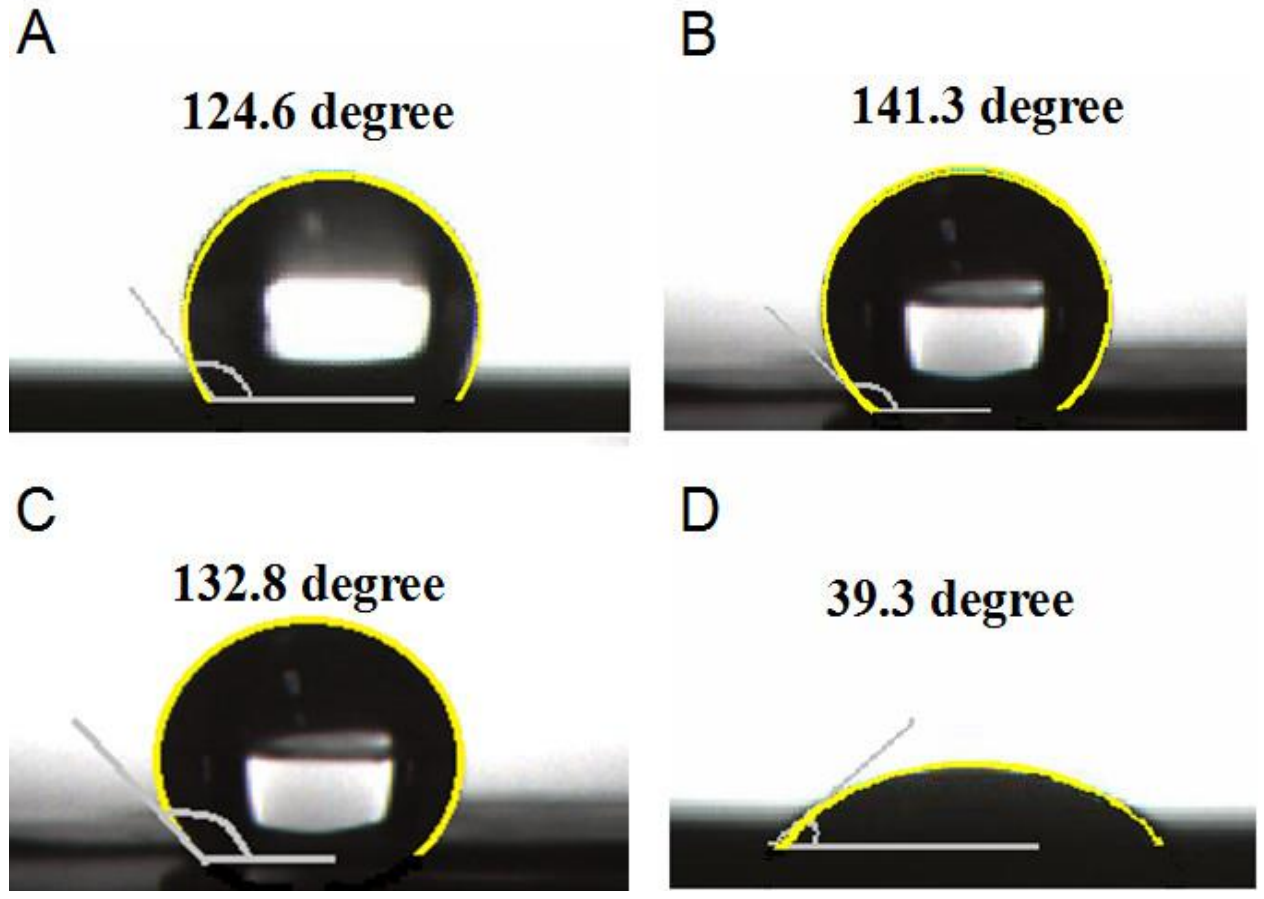

**Figure 2.**Optical images of a water droplet on TiNTs/AuNPs: (A) after modification with ODPA, (B) after modification with thiole NDM, (C) after loading with AMP, (D) after releasing AMP by visible light illumination for 30 min.

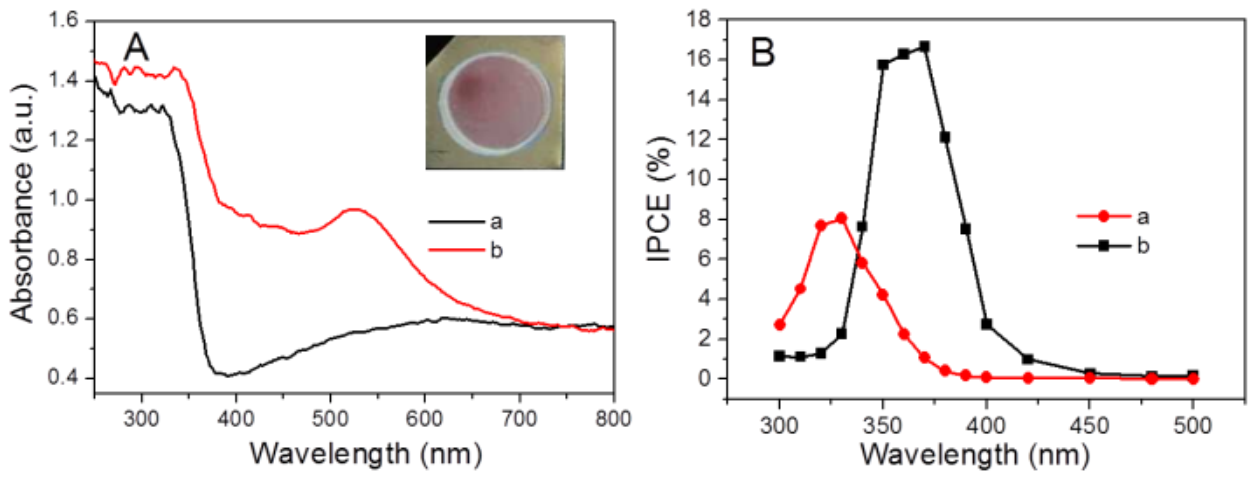

**Figure 3.** (A) UV-visible diffuse reflectance spectra of anatase $TiO_2$ nanotube arrays (curve a) and the amphiphilic $TiO_2$ nanotube arrays decorated with Au nanoparticles (curve b), inset: photograph of the amphiphilic $TiO_2$ nanotube arrays with Au nanoparticles decoration. (B) Incident photon-to-current conversion efficiency (IPCE) of ODPA decorated TiNTs (curve a) and ODPA decorated AuNPs/TiNTs (curve b)in an aqueous solution of 0.1 M $Na_2SO_4$ at an applied bias of +0.5 V *vs* saturated calomel electrode (SCE).

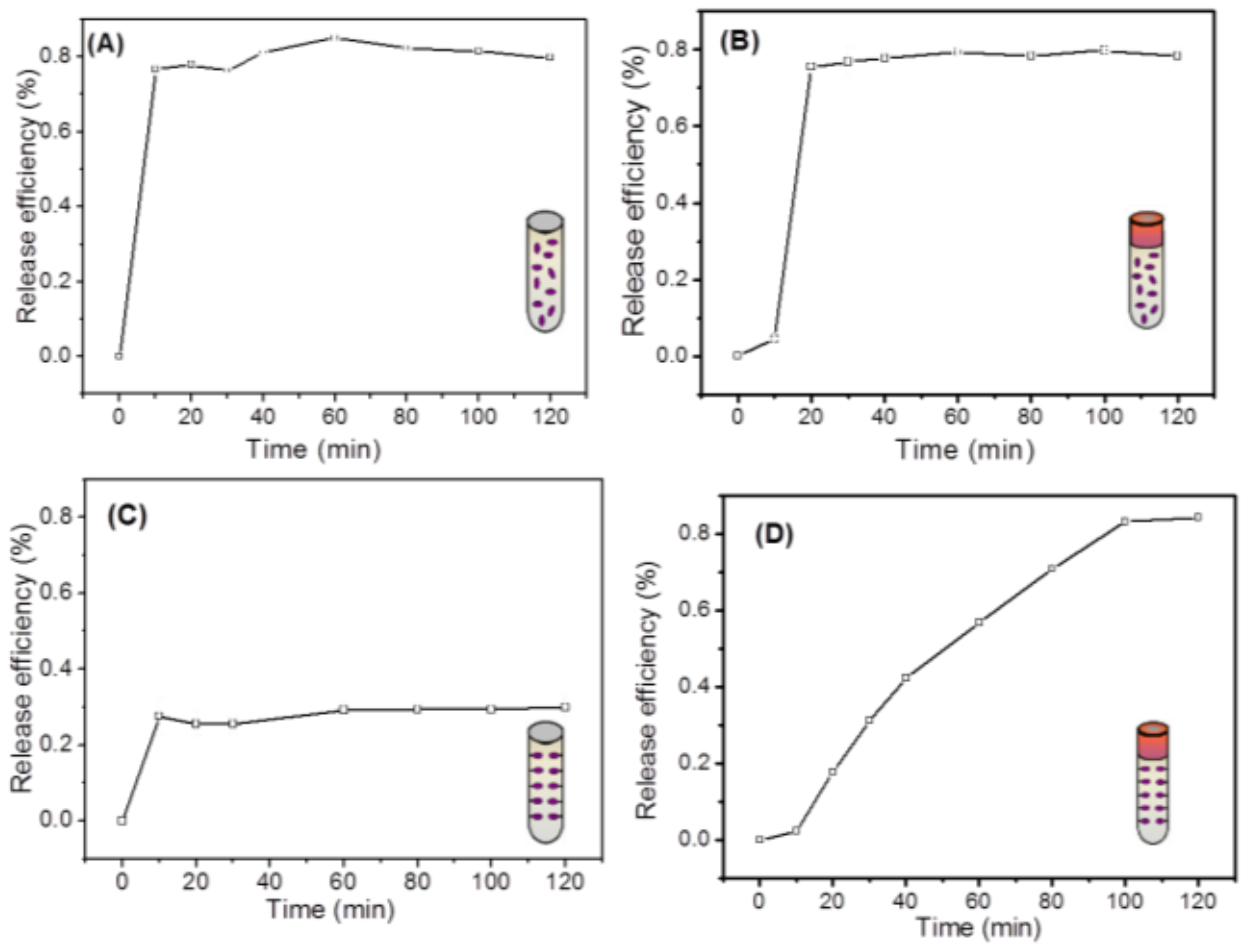

**Figure 4.** Release efficiency of AMP from the nanotubes. Four methods for drug loading using AMP as a hydrophilic model drug: (A) immersion without any $TiO_2$ surface modification, (B) immersion after ODPA modification in the upper nanotube layer (hydrophobic cap), (C) covalently attached HRP over the entire nanotube layers, (D) ODPA cap in the upper nanotube layer and covalently attached HRP in the lower nanotube layer.

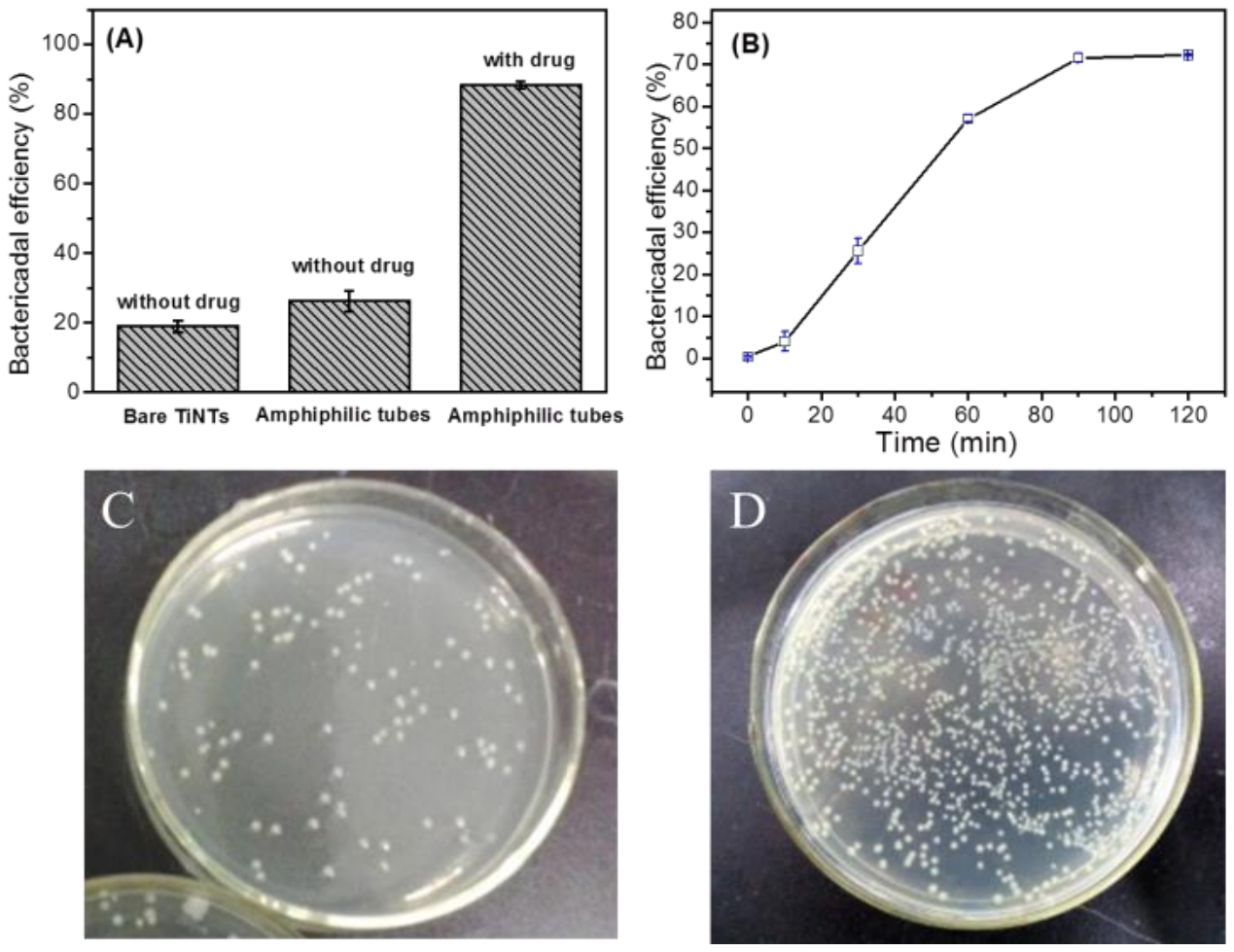

**Figure 5.**The bactericidal efficiency of (A) bare TiNTs, and amphiphilic TiNTs without and with drug loaded under visible-light irradiation. (B) AMP loaded amphiphilic TiNTs under visible-light illumination for different time. Optic images of bactericidal results from drug loaded amphiphilic TiNTs with (C) or without (D) treatment by visible-light irradiation.

# Supporting Information

## Visible Light Triggered Drug Release from $TiO_2$ Nanotube Arrays: A Novel Controllable Antibacterial Platform

Jingwen Xu,[a] Xuemei Zhou,[b] Zhida Gao,[a] Yan-Yan Song,[a]* and Patrik Schmuki[b]*

[a]College of Sciences, Northeastern University, Box 332, Shenyang 110004, China. E-mail: yysong@mail.neu.edu.cn. Tel.: +86-24-83687659.

[b]Department of Materials Science, *WW4-LKO*, University of Erlangen-Nuremberg, Martensstrasse 7, D-91058 Erlangen, Germany

## Experimental section

*1. Materials*

Ti sheets (0.1 mm thickness, 99.6% purity) were purchased from Advent Research (99.6% purity, England). Bovine Serum Albumin (BSA), octadecylphosphonic acid (ODPA), dodecanethiol (NDM), (3-Glycidyloxypropyl) trimethoxysilane (GPMS) were purchased from Sigma-Aldrich. $HAuCl_4$, glycerol, ammonium fluoride, ampicillin sodium (AMP) and other chemicals were purchased from Sinopharm Chemical Reagent Co. Ltd. and used without further purification. All aqueous solutions were prepared with deionized (DI) water.

*2. Preparation of amphiphilic $TiO_2$ nanotube arrays (TiNTs)*

Ti foils were first degreased by sonication in ethanol and DI water, and then dried in an $N_2$ stream. The double layer TiNTs were prepared using a concept we reported before but with some key modifications.[1] For this, cleaned Ti foils were anodized in an electrolyte of glycerol/water/NH4F at 15 V for 1 h, and then annealed at 450 °C in air. The resulting TiNTs were then incubated in Bovine Serum Albumin (BSA) solution (1 %) for 12 h, and further soaked in $HAuCl_4$ solution for another 4 h.[2] The $AuCl_4^-$ was absorbed by electrostatic interaction with the oppositely charged patches on the biomolecules. The resulting $AuCl_4^-$-BSA-TiNT hybrids were then exposed to UV light (Xenon lamp (300 W)) for 60 min. The photogenerated electrons in the TiNTs are driven to the surface of nanotubes under an interior field where they reduce the adsorbed $AuCl_4^-$ ions to $Au^0$. Simultaneously, the biotemplates are decomposed by the photogenerated holes ($h^+$). The AuNPs can be seen from the tube mouth to the tube (first layer) bottom. Then, a hydrophobic monolayer was decorated onto the tube walls by refluxing the AuNP decorated samples in 5 mM ODPA-toluene solution at 70 °C for 24 h. After ODPA modification, the samples were anodized again in an ethylene glycol electrolyte containing $NH_4F$ (0.135 M) at 30 V for 3 h. After cleaning by ethanol and DI water, the sample was incubated in the 2 mM NDM-toluene solution at room temperature for 24 h. These processes are illustrated in Scheme 1.

*3. Loading antibacerial drugs into the nanotubes*

Ampicillin sodium (AMP), a widely used antibacterial drug, was chosen as model drug to investigate the drug release kinetics, and the antibacterial effects towards *Escherichia coli* (*E. coli*). Grafting of AMP in the bottom layer of nanotubes was performed by silanization of the samples (Ti-OH) with GPMS.[3] Firstly, the samples (obtained in step *2*) were refluxed in 5 mM GPMS-toluene solution at 25 $^{o}$C for 12 h. Then AMP was covalently attached to the nanotubes by immersing the samples into 5 mL AMP–ethanol solution (10 mg/mL) at 4 °C for 24 h. The samples were then cleaned by DI water and dried with $N_2$. The samples should be kept in a dark box before use.

*4. Drug release*

The release of AMP was conducted in 5 mL DI water using a Xenon lamp (300 W) with a UV cutoff filter ($\lambda > 420$ nm) as visible-light source. The sample was attached to the bottom of a quartz cup containing 5 mL DI water. The lamp was located 20 cm from the reaction solution (50 mW cm$^{-2}$ illumination intensity). The drug concentration was measured by UV-vis spectrometry after reaction with ninhydrin to form colored products. The concentrations of drug solution before and after loading were respectively recorded as $C_o$ and $C_r$. After drug delivery, concentration of AMP in aqueous solution was recorded as $C_x$.

The release-efficiency of the drug was expressed as a percentage of released drugs, and can be calculated by the following equation:

$$\eta = \frac{m_x}{m_o - m_r} \qquad (1)$$

Where $\eta$ is the drug release efficiency, $m_o$ is the amount of drug in the original solution, $m_r$ is the remained drug in solution after loaded into nanotube arrays, and $m_x$ is the amount of released drug. The amount of drug can be calculated by measuring the drug concentration in a certain volume of solution.

*5. Antibacterial test*

*E. coli*, which is responsible for many infections in daily life, served as the model target microorganism for antibacterial tests. Before microbiological experiments, all glass ware were sterilized by autoclaving at 121 °C for 20 min. For each antibacterial experiment, 100 μL mixture of liquid Luria Broth (LB) substrate, 0.5% agar, and $10^7$ colony-forming unites per milliliter (CFU $mL^{-1}$) of *E. coli* were dropped on surfaces of drug-loaded samples. Samples were then exposed to visible-light (λ > 420 nm) for different time or in dark. After irradiation, the semi-solid bacteria solution was withdrawn and diluted serially with sterilized water to adjust the bacterial concentration to ensure the growing bacterial colonies were legible. In this concentration, 100 μL of the treated solution was spread on solid LB medium and the colonies were counted to determine the survival bacterial numbers after being constant temperature incubated at 37 °C for 24 h. To have a better comparison, bactericidal efficiency of anatase TiNTs and AuNPs decorated amphiphilic TiNTs (without AMP loading) were also treated using the same procedure. The bactericidal efficiency was determined by comparing the corresponding colony counts of control sample (sterilized glass sheet). All of the antibacterial experiments were repeated three times to give an average value.

*6.Analysis techniques*

The morphology of the nanotubes was characterized using a field-emission scanning electron microscope (Hitachi FE-SEM S4800, Japan). The UV-vis absorption spectra were measured on a spectrophotometer (Perkin–Elmer, Lambda XLS+, USA). X-ray photoelectron spectra (XPS) were recorded on a Perkin–Elmer Physical Electronics 5600 spectrometer. Transient photocurrent responses were acquired at an applied bias of +0.5 V under simulated AM 1.5 (100 $mW/cm^2$) illumination provided by a solar simulator (300 W Xenon lamp with a solar light filter, room temperature).

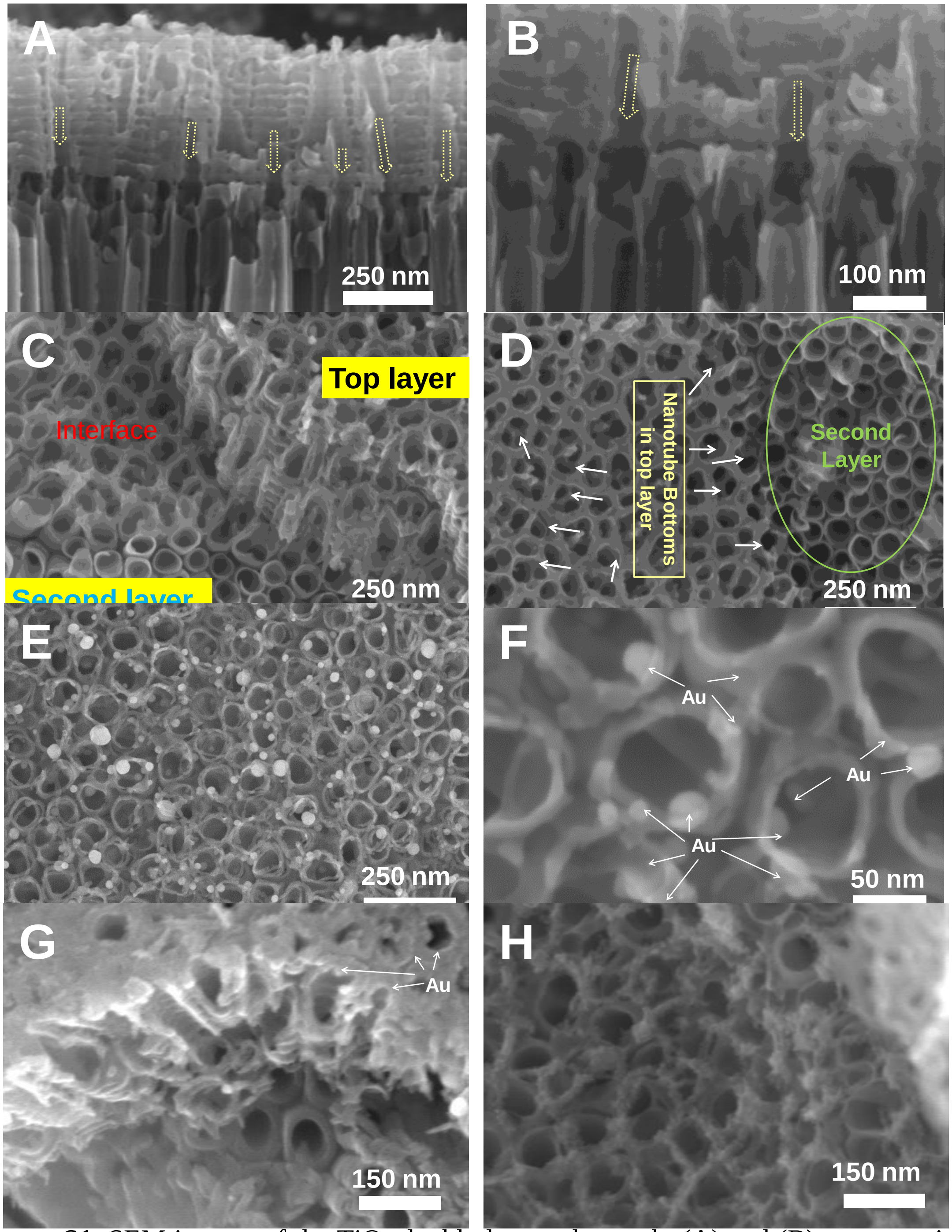

**Figure S1.** SEM images of the $TiO_2$ double-layer tube stack: (A) and (B) cross-sectional images and (C) and (D) interface between the upper and the lower $TiO_2$ nanotube layer. SEM images of the AuNPs decorated amphiphilic $TiO_2$ nanotube layers. Top-view images of the first nanotube layer (E) and (F), and cross-sectional images of the interface between the upper and the lower $TiO_2$ nanotube layers (G) and (H).

From SEM characterization of the cross-sectional images of the interface between the upper and the lower $TiO_2$ nanotube layers, it can be seen that the second (bottom) neat tube layer is grown underneath and interconnected with the first one. Clearly, the second anodization penetrates the bottom of the tubes grown during the first anodization.

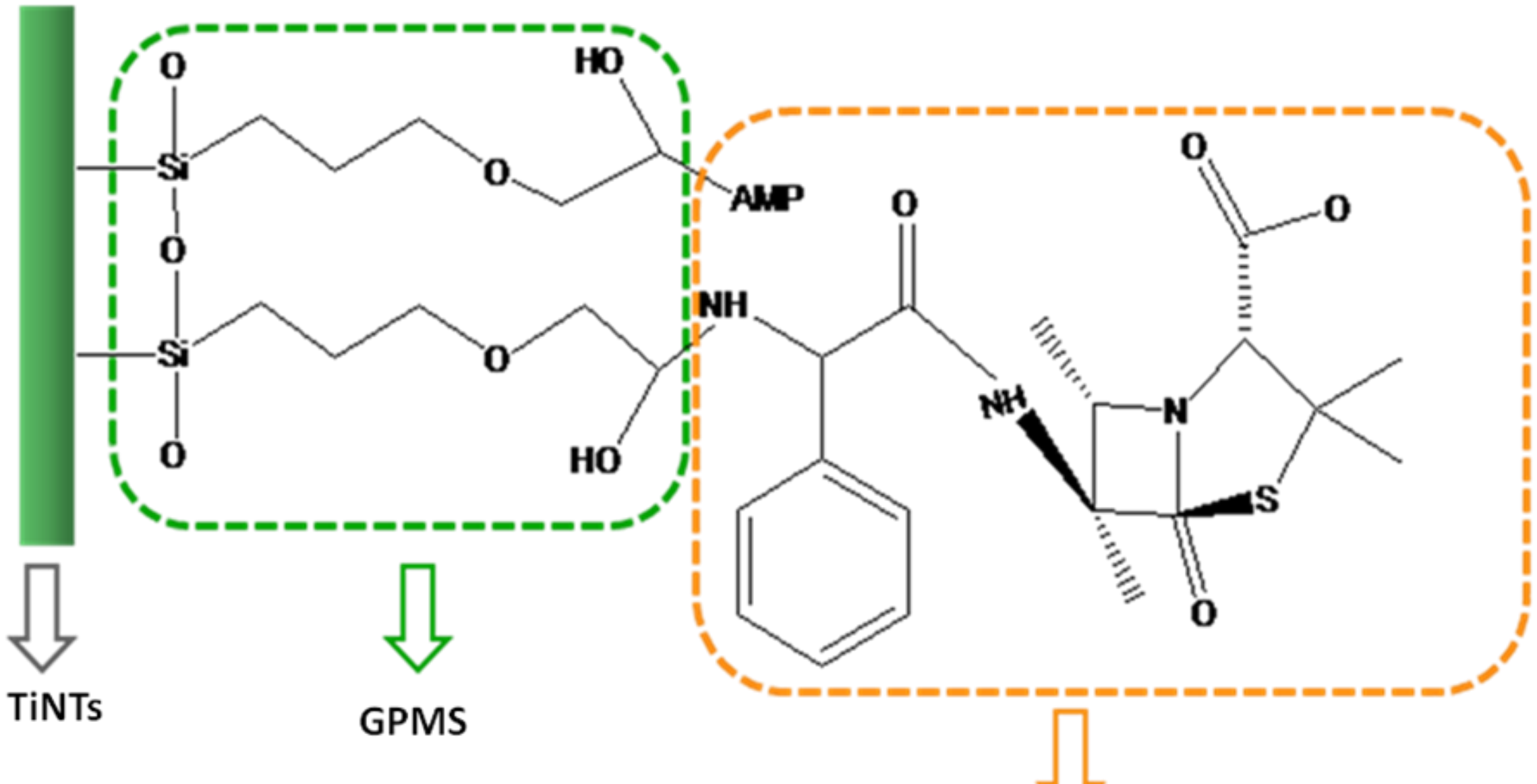

**Figure S2.** Schematic illustration of the grafting of AMP on $TiO_2$ naotubes (TiNTs) based on the silane linker GPMS.

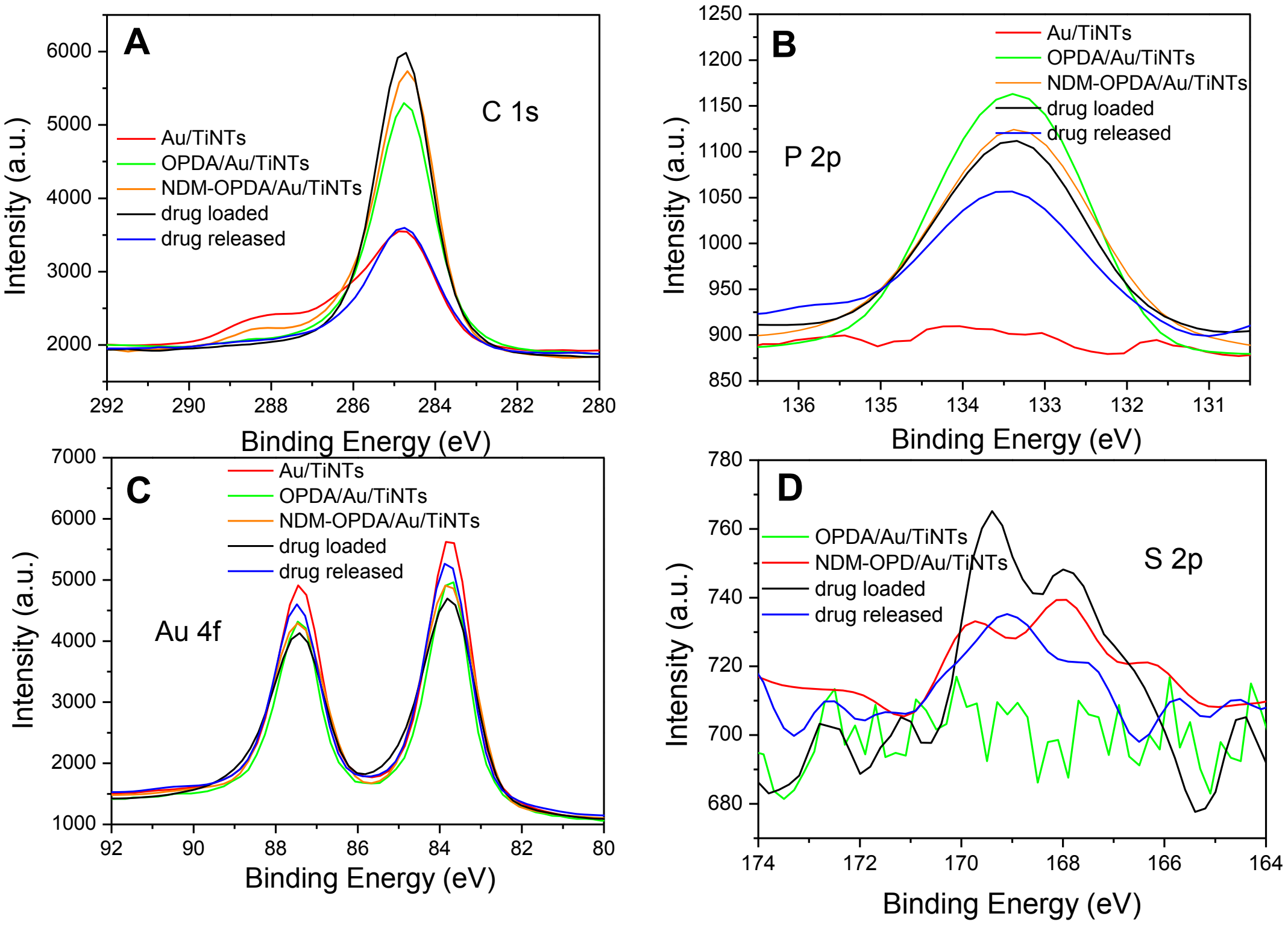

**Figure S3.** XPS characterization of Au grafting, amphiphilic nanotubes preparing, drug loading and drug release processes by visible light on $TiO_2$ nanotube layers.

To further investigate the processes of preparing amphiplilic nanotubes, as well as loading and release of the drug from nanotubes, XPS was employed to characterize the C, P, Au and S elements at different stages of the processes. Figure S3 shows XPS results recorded from the top surface of amphiphilic nanotube layer under identical XPS conditions. Figure S3A shows the intensity of C 1s peak of at 284.8 eV (C-C bonds) increases significantly after ODPA grafting and slightly after the NMD modification and drug loading. And an obvious decrease of intensity for C 1s after visible-light irradiation can be seen, the intensity almost returns to the original amount (Au/TiNTs). This indicates the aliphatic chains were cut off during visible light illumination. The release of drug by visible light is also approved from the P 2p peak (Figure S3B). The intensity of P 2p exhibits a partial decrease after the visible light irradiation. The appearance of higher binding energy component (~ 136.0 eV) is noticed after exposure to visible-light irradiation, which may be associated with the phosphorus in a higher oxidation state, possibly corresponding to a phosphate species rather than a phosphonate after undergoing the photocatalytic reaction.[4] The Au 4f peaks (Figure S3C) are visible during the whole process, suggesting that AuNPs grafted on the first layer are stable against the second anodization and drug release steps. However, a slight increase of intensity of the Au 4f peaks can be noticed after drug release – this may be due to monolayers that cover the metallic Au are cut off by visible light. The intensities of S 2p signals from NDM (Figure S3D) are relatively lower compared with other elements. One possibility is that the absorption sites on Au surfaces are occupied first by ODPA,[5] so that NDM can merely react with either the rest of Au surfaces or the defects formed by the second anodization step. Therefore, there would be a decrease of the adsorption of silane and AMP during the drug loading process.

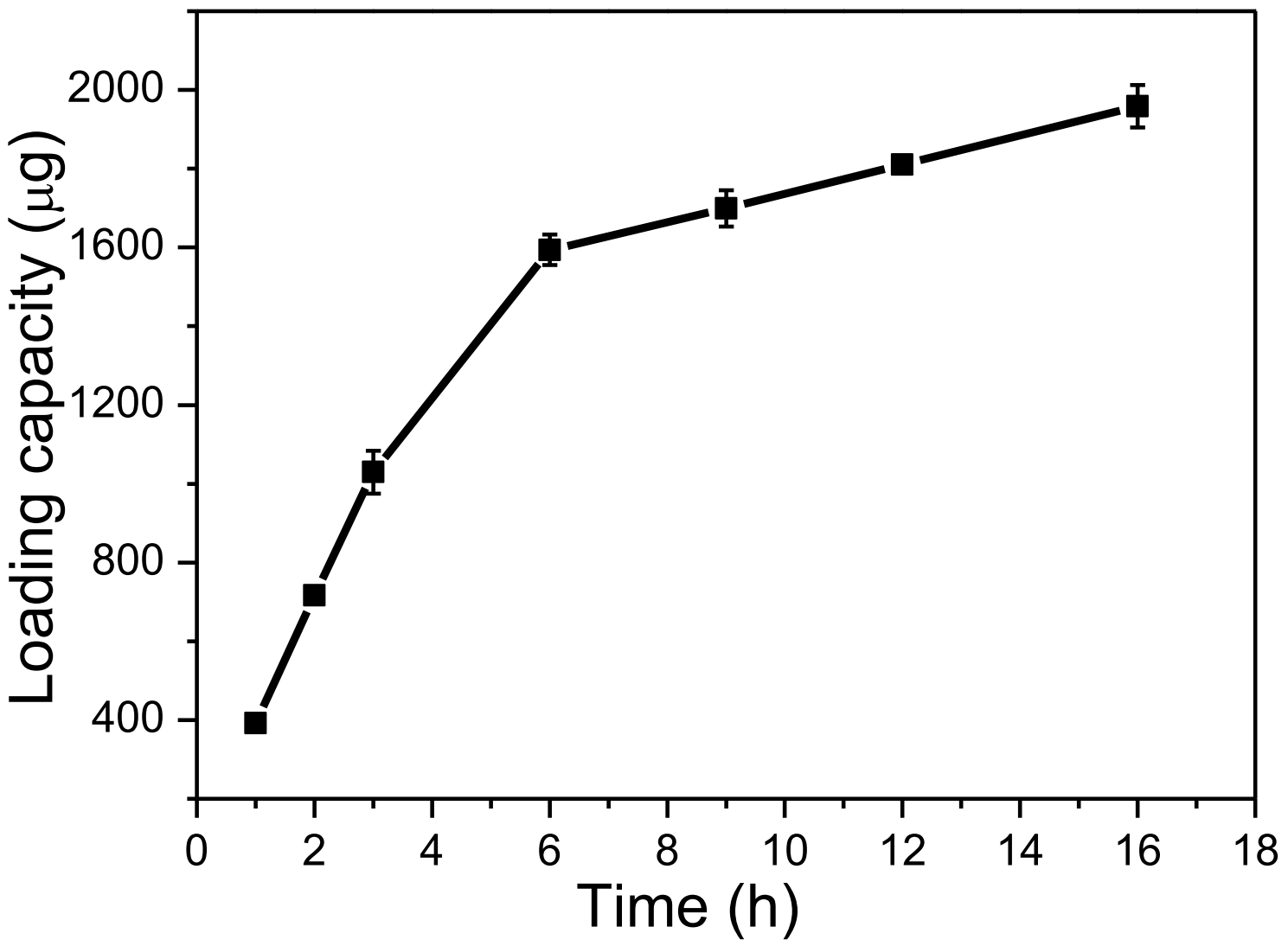

**Figure S4.** Influence of anodization time on loading capacity of photoinduced drug-release TiNTs.

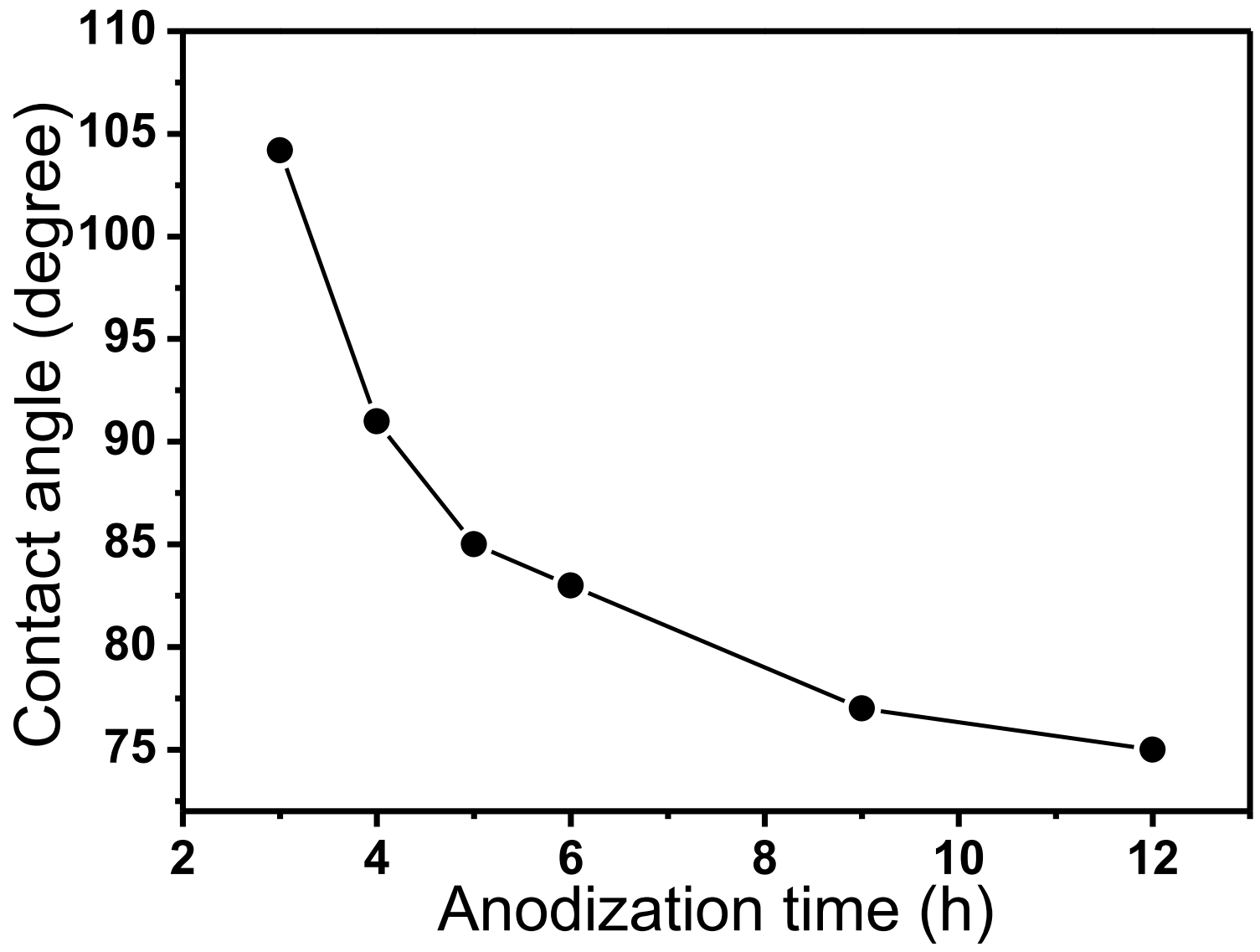

**Figure S5.** Influence of the second anodization time on the contact angle of water droplets on as-formed amphiphilic sample surface.

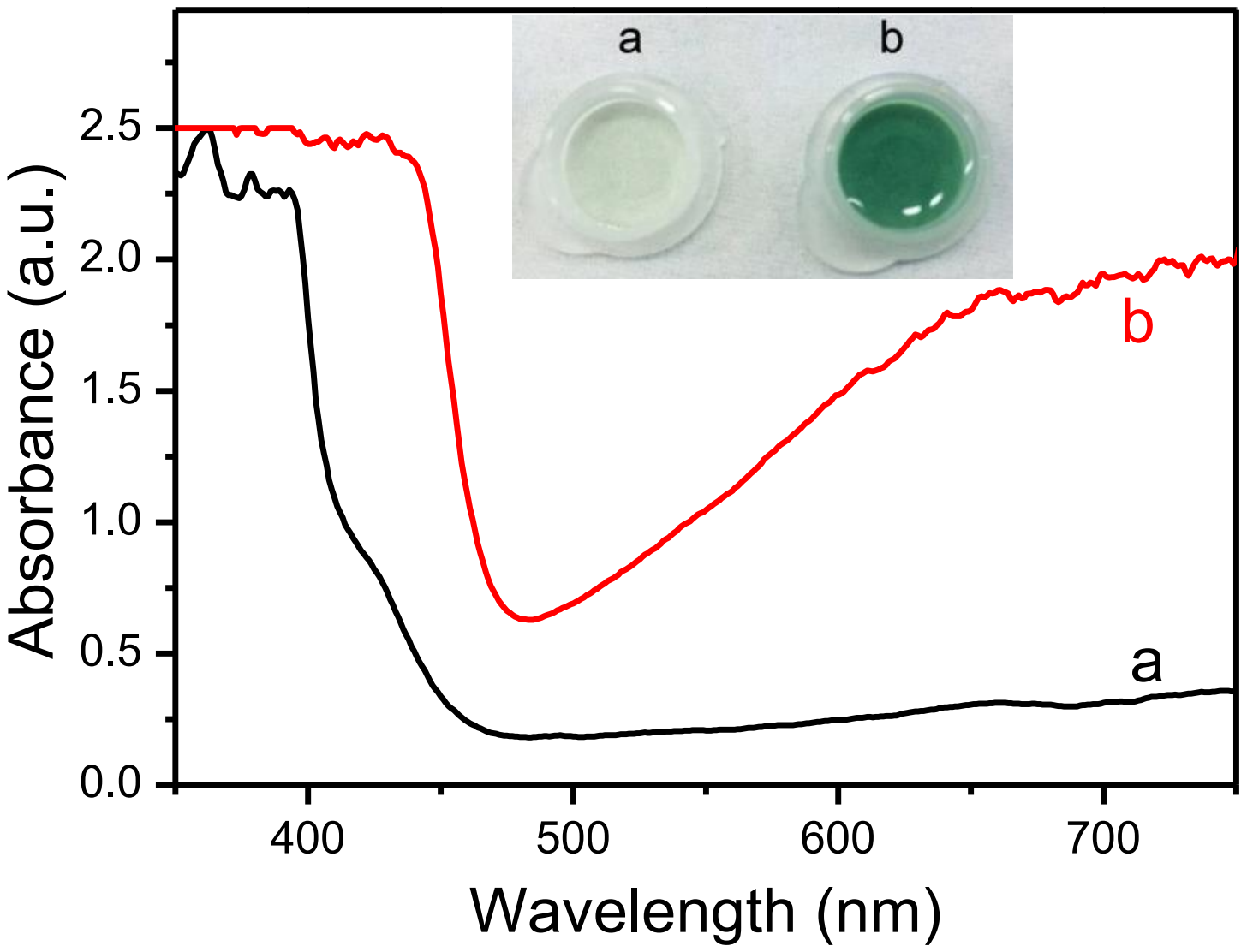

**Figure S6.** Demonstration of the formation of $H_2O_2$ during drug release process by visible light. After 10 min visible-light illumination solution in vicinity of the AuNPs decorated amphiphilic nanotubes is collected by a micropipette and transferred to a 0.4 mL of PBS solution (pH 6.0) containing 0.01 mg/mL HRP and 0.05 M ABTS as the indicator substrate. UV-vis spectra show the solution containing ABTS and HRP before (curve a) and after (curve b) the color reaction.[6] Insert：the optical images of the solution corresponding to curve a and curve b. From the color reaction, it is clear that the formation of $H_2O_2$ during the visible-light induced drug release process.

As a typical electron acceptor, $O_2$ takes photoexcited electrons from the gold nanoparticles directly so as to reduce the photoanodic current, and induces $O_2^{\bullet}$ in solution. Since $O_2^{\bullet}$ can react with $H_2O$ to form $H_2O_2$, the color reaction of HRP and 2, 2'-azino-bis (3-ethylbenzothiazoline-6-sulforic acid) can be used to demonstrate the existence of $H_2O_2$. As shown in Figure S6A, $H_2O_2$ is detected in the solution after Au-TiNTs exposed to visible light ($\lambda > 420$ nm) illumination for 10 min. This result supports the reduction of oxygen, since $H_2O_2$ is one of the possible products of oxygen reduction.

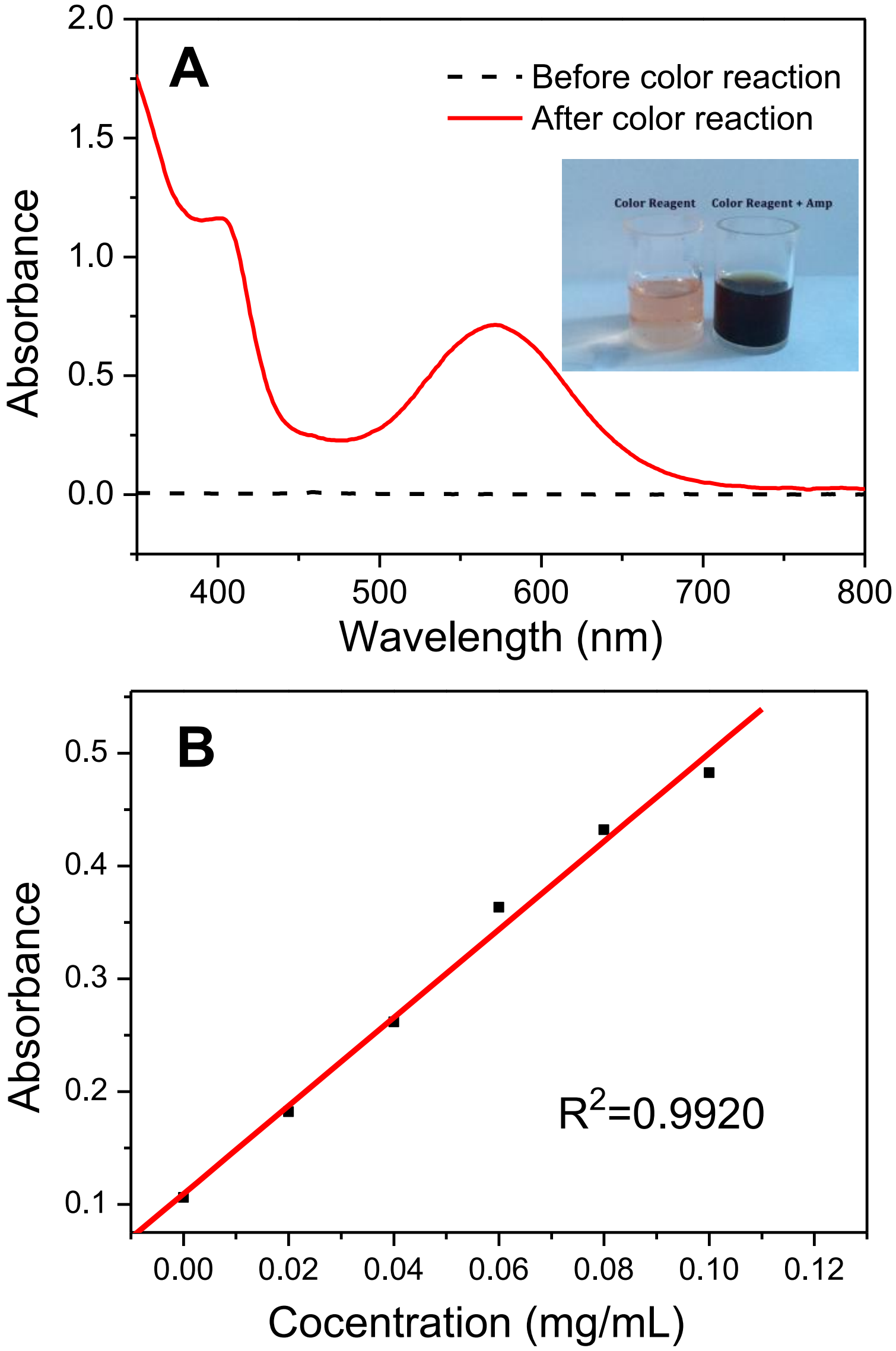

**Figure S7.** (A) UV-vis spectra of AMP before and after chromogenic reactionwith ninhydrin (insert: the optical images of AMP before and after chromogenic reaction). (B) Corresponding calibration curve of AMP concentration.

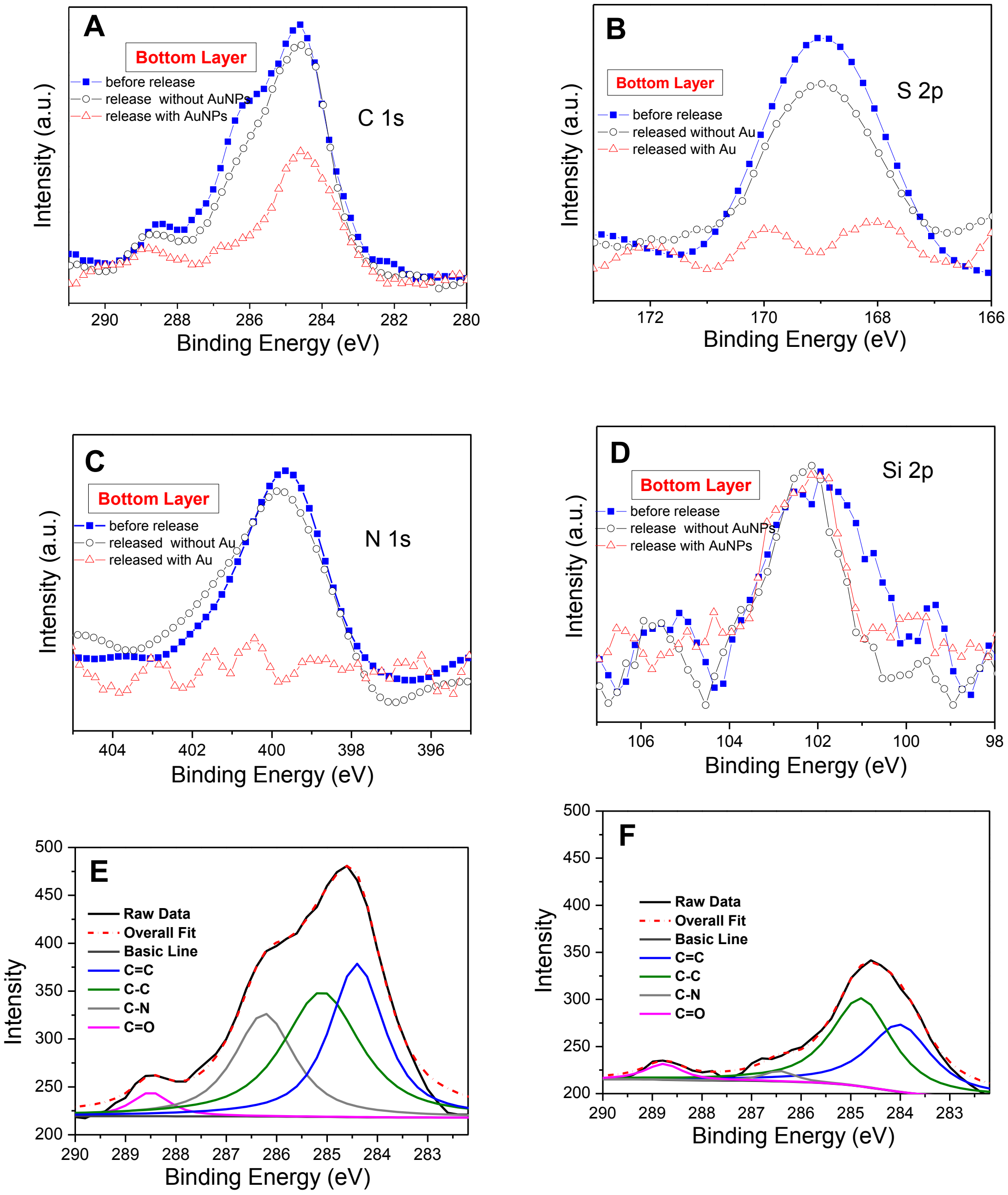

**Figure S8.** XPS spectra of lower nanotube layer to show the influence of AuNPs towards the visible-light induced drug release.XPS spectra show (A) C 1s, (B) S 2p, and (C) N 1s signals in the bottom nanotube layer.

Figure S8 shows the XPS peaks of C, S, N and Si elements obtained from the bottom layer of the drug system (illustrated as last step in Scheme 1 in the main text). The top tube layer was removed (as illustrated in Figure S9). The results confirm the drug molecules grafted in the bottom layer can be released successfully by visible-light irradiation when the Au NPs are successfully attached on the top layer (red triangles curves) - the intensity of S 2p and N 1s peaks decreases, and a significant peak shift in C 1s spectra can be observed. However, if this Au decorated top layer is removed before the visible light irradiation (black cycles), the drug molecules couldn't be released from $TiO_2$ nanotubes by visible-light irradiation, i.e. no obvious signal decrease in XPS spectra can be seen. Only a few physisorbed drug molecules are released from the system. These data demonstrate that the AuNPs in the top layer are the key to induce visible-light triggered drug release. Figure S8E and Figure S8F are the high-resolution C 1s peaks before and after drug release, respectively. The peaks for C=C, C-N, and C=O peaks (from AMP, as shown in Figure S2) are decreased after the drug release.

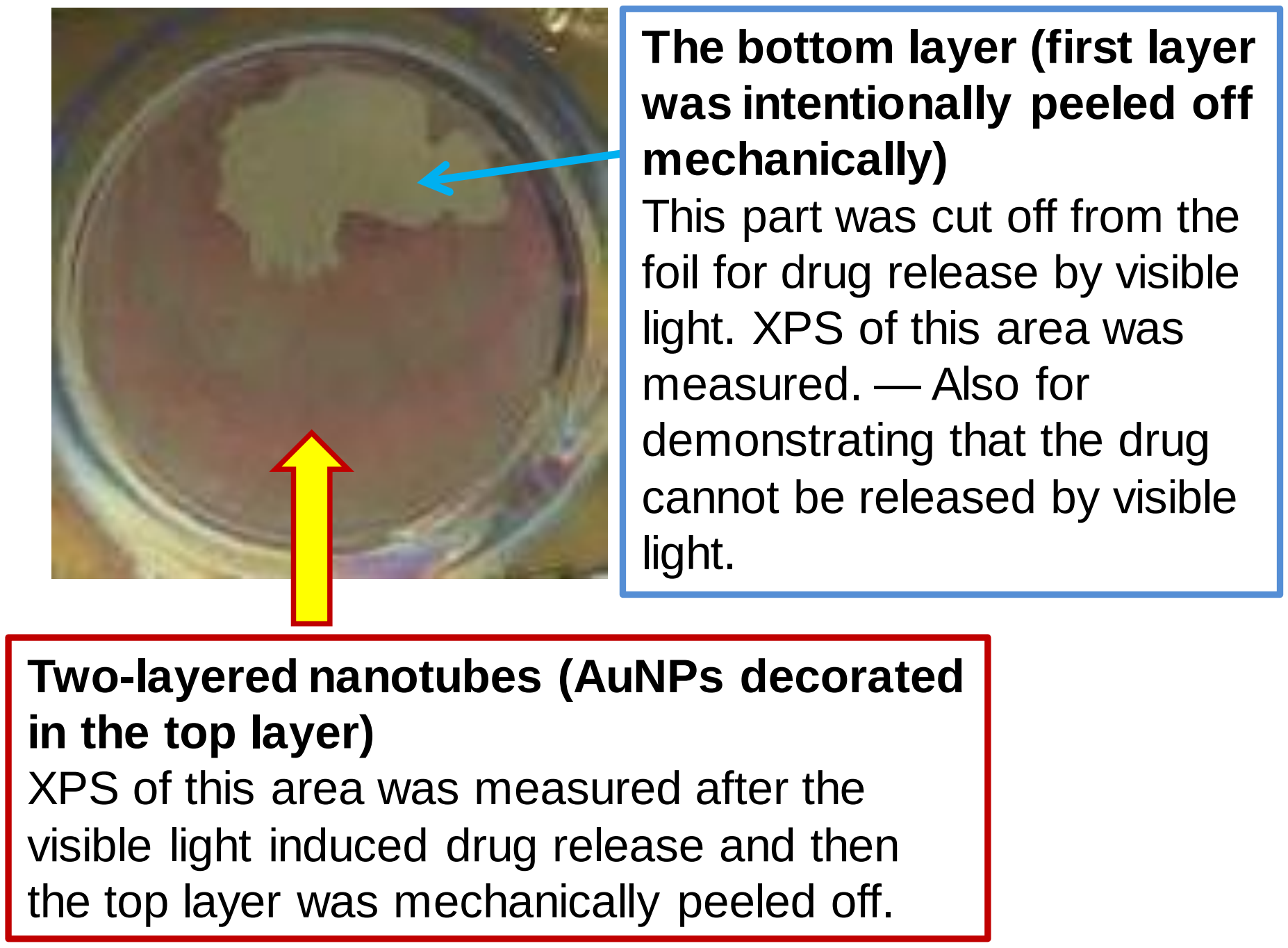

**Figure S9.** Photograph of the $TiO_2$ nanotube arrays used in Figure S8.